\documentclass[traditabstrac]{aa} 

\newcommand{\micron}{$\mu$m}

\usepackage{graphicx}
\usepackage{txfonts}

\usepackage{graphicx}
\usepackage{txfonts}

\begin{document}

   \title{ALMA continuum observations of the protoplanetary disk AS 209}
 \subtitle{Evidence of multiple gaps opened by a single planet}

   \author{D. Fedele
          \inst{\ref{inst_inaf}}
          \and
          M. Tazzari\inst{\ref{inst_cam}}
          \and
          R. Booth\inst{\ref{inst_cam}}
          \and
          L. Testi\inst{\ref{inst_eso}}
          \and
          C.~J. Clarke\inst{\ref{inst_cam}}
        	\and	
	\\
          I. Pascucci\inst{\ref{inst_lpl}}
	\and
          A. Kospal\inst{\ref{inst_kon}}
          \and
          D. Semenov\inst{\ref{inst_mpia}}
          \and
          S. Bruderer
          \and
          Th. Henning\inst{\ref{inst_mpia}}
          \and
          R. Teague\inst{\ref{inst_mpia}}
          }

\institute{
INAF-Osservatorio Astrofisico di Arcetri, L.go E. Fermi 5, I-50125 Firenze, Italy\label{inst_inaf}\\
\and
Institute of Astronomy, University of Cambridge, Madingley Road, Cambridge CB3 0HA, UK\label{inst_cam}\\
\and
European Southern Observatory, Karl-Schwarzschild-Strasse 2, 85748, Garching bei Muenchen, Germany\label{inst_eso}\\
\and
Lunar and Planetary Laboratory, The University of Arizona, Tucson, AZ 85721, USA\label{inst_lpl}\\
\and
Konkoly Observatory, Research Centre for Astronomy and Earth Sciences, Hungarian Academy of Sciences, Konkoly-Thege Mikl\'os \'ut 15-17, 1121 Budapest, Hungary\label{inst_kon}
\and
Max Planck Institute for Astronomy, K\"onigstuhl 17, 69117, Heidelberg, Germany\label{inst_mpia}\\
}
\date{Received ...; accepted ...}
 
  \abstract
   {The paper presents new high angular resolution ALMA 1.3\,mm dust continuum observations of the protoplanetary system AS 209 in the Ophiuchus star forming region.
    The dust continuum emission is characterized by a main central core and two prominent rings at $r = 75\,$au and $r = 130\,$au intervaled by two
    gaps at  at $r = 62\,$au and $r = 103\,$au. The two gaps have different widths and depths, with the inner one being narrower and shallower.
     We determined the surface density of the millimeter dust grains using the 3D radiative transfer disk code \textsc{dali}. According to our fiducial model the 
    inner gap is partially filled with millimeter grains while the outer gap is largely devoid of dust.
    The inferred surface density is compared to 3D hydrodynamical simulations (FARGO-3D) of planet-disk 
    interaction. The outer dust gap is consistent with the presence of a giant planet ($M_{\rm planet} \sim 0.8\,M_{\rm Staturn}$); the planet is responsible for the 
    gap opening and for the pile-up of dust at the outer edge of the planet orbit. The simulations also show that the same planet can give origin to the inner gap 
    at $r = 62\,$au. 
    The relative position of the two dust gaps is close to the 2:1 resonance and we have investigated the possibility of a second planet inside the 
    inner gap. The resulting surface density (including location, width and depth of the two dust gaps) are in agreement with the observations.  The 
    properties of the inner gap pose a strong constraint to the mass of the inner planet ($M_{\rm planet} < 0.1\,M_{\rm J}$). In both scenarios (single 
    or pair of planets), the hydrodynamical simulations suggest a very low disk viscosity ($\alpha < 10^{-4}$).  Given the young age of the system 
    (0.5 - 1\,Myr), this result implies that the formation of giant planets occurs on a timescale of $\lesssim$ 1\,Myr.}

   \keywords{giant planet formation -- T Tauri }

   \maketitle
%

\section{Introduction}
Axisymmetric gaps and rings such as those seen in the protoplanetary disks around HL Tau, TW Hya, HD163296, HD 169142, AA Tau 
\citep[e.g.][]{alma15, Andrews16, Isella16, Fedele17, Loomis17} can now regularly be unveiled by the extremely high resolution available 
with ALMA.  
The formation of gaps and rings in disks can be due to several mechanisms such as: planet formation \citep[e.g.,][]{Papaloizou84}; 
magneto-rotational instability \citep{Flock15}; condensation fronts \citep{Zhang15}; dust sintering \citep{Okuzumi16}; photoevaporation 
\citep{Ercolano17}. 

\smallskip
\noindent
The rings observed by ALMA show that the millimeter dust grains are radially confined  regions in which inward radial migration is  slowed down or 
stopped and so may be key to explaining the retention of large grains in disks on long ($2-3$\,Myr) timescales (irrespective of the mechanism 
producing such ``dust traps''). 
In addition, dust traps provide the ideal environment in which to observationally constrain models of grain growth because -- in contrast to 
other regions of the disk -- they are regions in which radial drift is relatively unimportant, allowing  dust to grow {\it in situ} \citep{Pinilla12}. 
Given a measure of the local dust density, the timescale for dust growth to a given size is readily obtained from grain growth models 
\citep[e.g.,][]{Birnstiel12}.
Empirical measurement of the maximum grain size in traps would thus provide the cleanest test of the various assumptions (sticking probability, 
turbulent velocity, fragmentation threshold) that enter these models.

\begin{figure*}[!t]
\centering
\includegraphics[width=18cm]{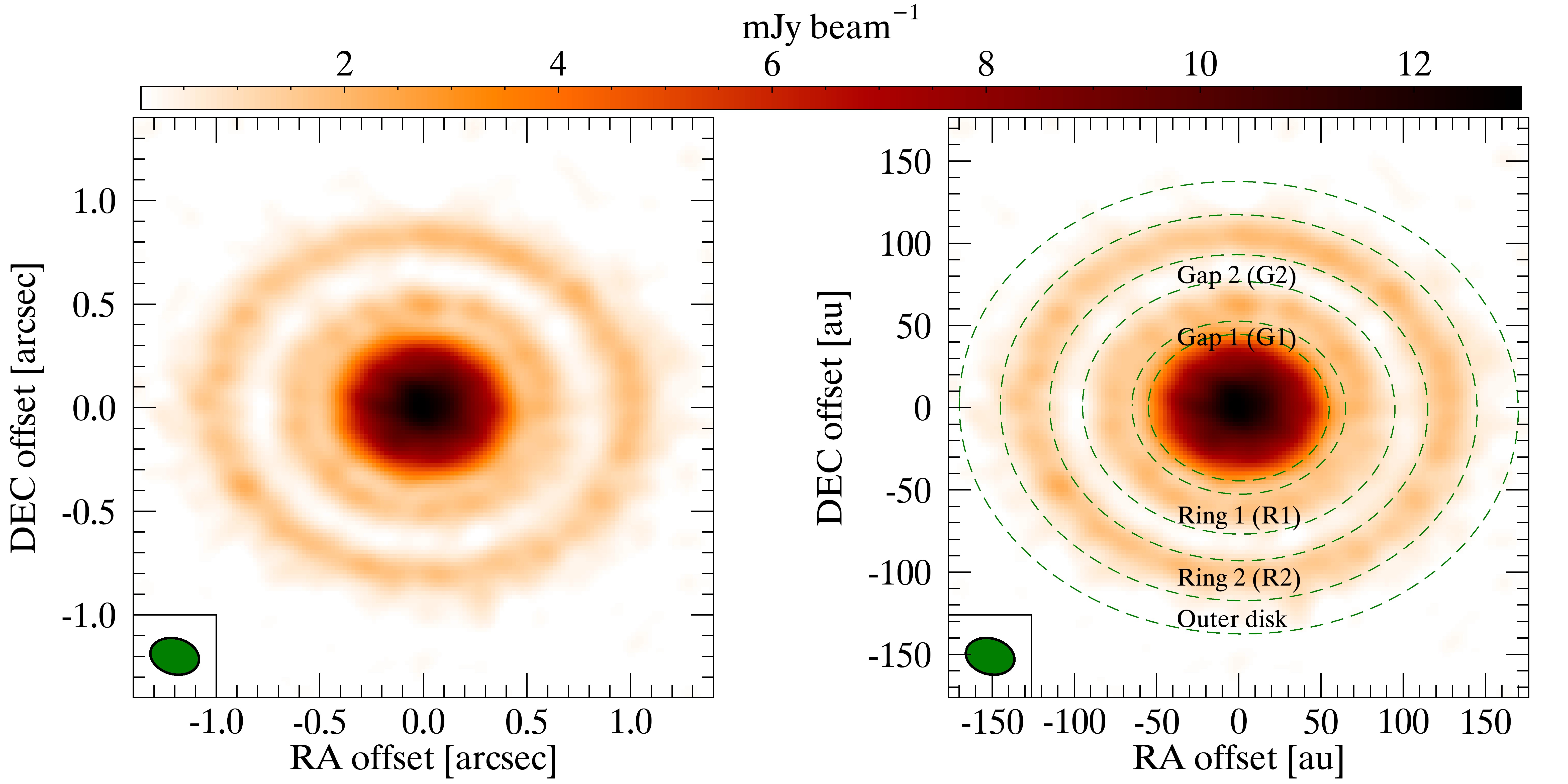}
\caption{ALMA 1.3\,mm dust continuum image (uniform weighting). The main substructures are highlighted in the right panel.}\label{fig:continuum}
\end{figure*}

\smallskip
\noindent
This paper presents new ALMA 1.3\,mm continuum observations of the T Tauri disk AS~209 where axisymmetric gaps and rings 
are detected.
AS~209 ($M_{\star} = 0.9\,M_{\odot}$, spectral type K5, $L_{\star} = 1.5\,L_{\odot}$, \citealt{Tazzari16}) is part of the young 
(age $\sim 0.5 - 1.0\,$Myr \citealt{Natta06}) Ophiuchus star forming region at a distance of 126\,pc from the Sun \citep{Brown16}. 
Multi-frequency continuum observations revealed optically thin emission at millimeter wavelenghts beyond a few 10s of ~au
from the star \citep{Perez12, Tazzari16}. \citet{Huang16} found evidence of an extended gas emission (C$^{18}$O) speculating that it is due
to external CO desorption in the outer disk. Interestingly, \citet{Huang17} noticed the presence of a dark lane in the dust 1.1\,mm continuum emission.

\smallskip
\noindent
The structure of the paper is the following: observations and data reduction are presented in section~\ref{sec:obs} and the results are 
discussed in section~\ref{sec:results}. The data analysis is described in Section~\ref{sec:analysis}. Section~\ref{sec:hydro} provides a comparison
to hydrodynamical simulations. Discussion and conclusion are reported in sec.~\ref{sec:conclusion}.

\section{Observations and Data reduction}\label{sec:obs}
The ALMA observations of AS 209 (J2000: R.A. = 16$^{\rm{h}}$49$^{\rm{m}}$15.296$^{\rm{s}}$, 
DEC = --14$^\circ$22$\arcmin$09.02$\arcsec$) have been performed on 2016 September 22 (with 38 antennas) and 26 (41
antennas) in band 6 (211--275\,GHz) as part of the project ID 2015.1.00486.S (PI: D. Fedele). The correlator 
setup includes a broad (2\,GHz bandwidth) spectral window centered at 230\,GHz.

\smallskip
\noindent
Visibilities were taken in two execution blocks with a 6.05s integration time per visibility totalling 40 minutes, per block, on-source. System 
temperatures were between $80-145$\,K. Weather conditions on the dates of observation gave an average precipitable
water vapour of 2.2 and 2.3\,mm, respectively. Calibration was done with $J1517-2422$ as bandpass calibrator, $J1733-1304$ as 
phase and flux the flux calibrator.  
 The visibilities were subsequently time binned to 60s integration times per visibility for self-calibration, imaging, and analysis.
Self-calibration was performed using the 233 GHz continuum TDM spectral window with DA41 as the reference antenna. 

\smallskip
\noindent
The continuum image was created using \textsc{casa.clean} (\textsc{casa} version 4.7); after trying different weighting schemes, we opted
for a uniform weight which yields a synthesised beam of 0\farcs19 $\times$ 0\farcs14 (PA = 75.5$^{\circ}$). The 
peak flux is 13\,mJy\,beam$^{-1}$ and the r.m.s. is 0.1\,mJy\,beam$^{-1}$.

\section{Results}\label{sec:results}
\begin{figure}[!t]
\centering
\includegraphics[width=\columnwidth]{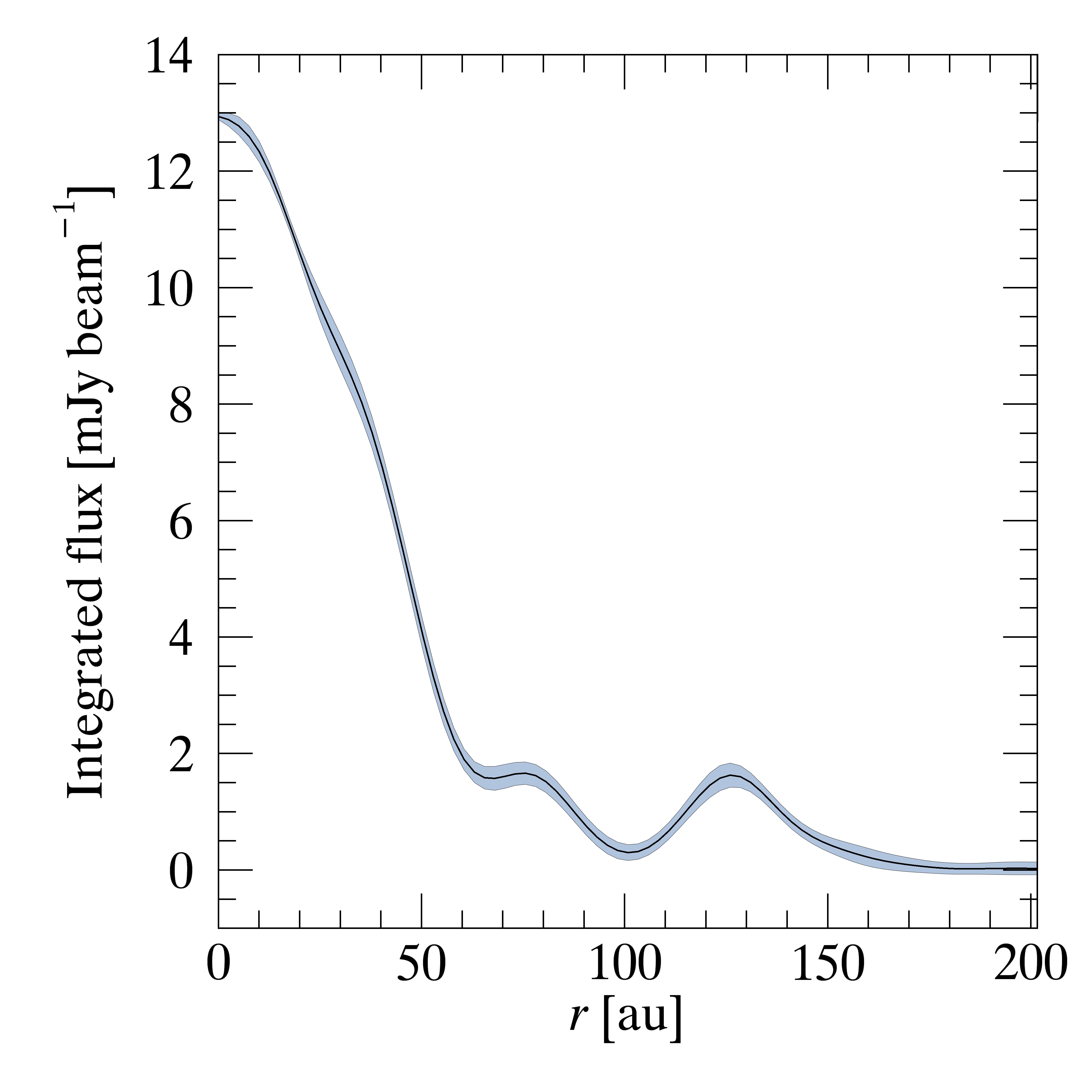}
\caption{Radial intensity profile of the 1.3,\,mm dust continuum emission. The profile is azimuthally averaged after deprojecting for the disk 
inclination ($i = 36^{\circ}$, Sect.~\ref{sec:results}). The black line shows the mean profile while the shadowed regions indicate the standard
deviation along the azimuth angle.}\label{fig:radial}
\end{figure}

The ALMA 1.3\,mm dust continuum image is shown in Fig.~\ref{fig:continuum}: the continuum emission is characterized
by a bright central emission and two weaker dust rings peaking at $\sim 75\,$au and $130\,$au, respectively. The two rings 
have a similar peak flux ($\sim 2\,$mJy). The rings are intervaled by two narrow gaps. The two gaps have different widths and depths.
The radial intensity profile shows a kink around $20-30\,$au which may be the signature of another (spatially unresolved) dust gap.
Finally, the continuum flux does not drop to zero at the edge of the outer ring as there is a tenuous emission extending out to $\sim 170-180\,$au. 
The different disk substructures are clearly visible in the radial intensity profile shown in Fig.~\ref{fig:radial}. 

\subsection{Characterization of the brightness profile}
In this section we present here the fit of the observed visibilities, which provides an initial characterization of the disk surface brightness useful for the 
detailed physical modelling carried out in Section~\ref{sec:analysis}. We assume an axisymmetric brigthness profile defined as follows:

\begin{equation}
I(R)=\delta(R) \ I_0 \left( \frac{R}{R_c} \right)^{\phi_1} 
\exp \left[ -\left( \frac{R}{R_c} \right)^{\phi_2} \right],
\end{equation}

where $I_0$ is a normalization, $R_c$ is a scale length and $\delta(R)$ is a scaling factor (by definition $\delta(R)>0$) parametrized as:

\begin{equation}
\delta(R)=
\begin{cases}
\delta_{\rm G1}		& \qquad \text{for  R $\in$ [$R_{\rm G1} - hw_{\rm G1}, R_{\rm G1} + hw_{\rm G1}$]}\\
\delta_{\rm R1}       	& \qquad \text{for  R $\in$ [$R_{\rm G1} + hw_{\rm G1}, R_{\rm G2} - hw_{\rm G2}$]}\\
\delta_{\rm G2}       	& \qquad \text{for  R $\in$ [$R_{\rm G2} - hw_{\rm G2}, R_{\rm G2} + hw_{\rm G2}$]}\\
\delta_{\rm R2}       	& \qquad \text{for  R $\in$ [$R_{\rm G2} + hw_{\rm G2}, R_{\rm R2, out} $] } \\
\delta_{\rm out}       	& \qquad \text{for  R $\geq  R_{\rm R2, out}$}\\
1                                & \qquad \text{otherwise}
\end{cases}
\end{equation}\label{eq:delta}

\noindent
where $R_{\rm G}$ and $hw_{\rm G}$ are the center and half width of the dust gaps, respectively. 
The choice of this particular brightness profile serves as a simple realization 
of an ``unperturbed'' profile (an exponentially tapered power law), characterized by 
a few radial regions that can depart from it either due to an excess ($\delta>1$) 
or a lack ($\delta<1$) of emission. Following the evidence emerging from 
the synthesized image (Fig.~\ref{fig:continuum}), we allow for two rings, two gaps, and an outer disk region, 
following nomenclature in Fig.~\ref{fig:continuum}. In this framework, a gap in the disk is naturally modelled with $\delta<1$.

\begin{table}[t!]
\caption{Parameter space explored by the Markov chains and best-fit values}
\begin{center}
\begin{tabular}{lrrr}
\hline
\hline
Parameter   &         Min &     Max   &Best-fit \\
\hline
         I$_0$ [mJy/beam] &    0 & 100 & 	 7.4 	$\pm$ 0.2		\\
         $R_c$ [au] 	&     20   	& 150	&   	80 	$\pm$ 1		\\
    $\phi_1$          	&     -4 	& 4  	&   	 -0.24 	$\pm$ 0.01	\\
   $\phi_2$          	&     -4 	& 4  	&   	2.19 	$\pm$ 0.02	\\
      $R_{G1}$ [au] 	&     0 	& 80 	&   	61.7 	$\pm$ 0.5		\\
      $hw_{G1}$ [au] 	&     0 	& 30 	&   	 8.0 	$\pm$ 0.2		\\
 $\delta_{G1}$          &     0 	& 1  	&   	 0.03 $\pm$ 0.005	\\
 $\delta_{R1}$        	&     0 	& 3  	&   	0.80 	$\pm$ 0.02		\\
      $R_{G2}$ [au] 	&     80 	& 110	&   	103.2 $\pm$ 0.4		\\
      $hw_{G2}$ [au] 	&     0		& 30 	&   	15.6 	$\pm$ 0.2		\\
 $\delta_{G2}$          &     0		& 1  	&   	 0.025 $\pm$ 0.005	\\
 $\delta_{R2}$          &     0		& 20 	&   	 4.8  $\pm$ 0.1	\\
 $R_{R2, out}$ [au] 	&     130 	& 180	&   	139.8  $\pm$ 0.8		\\
$\delta_{out}$          &     0 	& 2  	&   	 1.95 	$\pm$ 0.03	\\
           $i$ [$^\circ$]&     0 	& 90 	&   	 35.3 	$\pm$ 0.8		\\
          $PA$ [$^\circ$]&     0 	& 180	&   	 86.0 	$\pm$ 0.7		\\
\hline
\end{tabular}
\end{center}
\label{tab:mcmc}
\end{table}%

\smallskip
\noindent
We perform the fit of the visibilities with a Bayesian approach using the Monte Carlo Markov chains ensemble sampler  
\citep{Goodman10} implemented in the \textsc{emcee} package \citep{Foreman13}. We assume flat priors for all the  
free parameters. Table~\ref{tab:mcmc} reports the ranges explored for each parameter.

\begin{figure}[!t]
\centering
\includegraphics[width=\columnwidth]{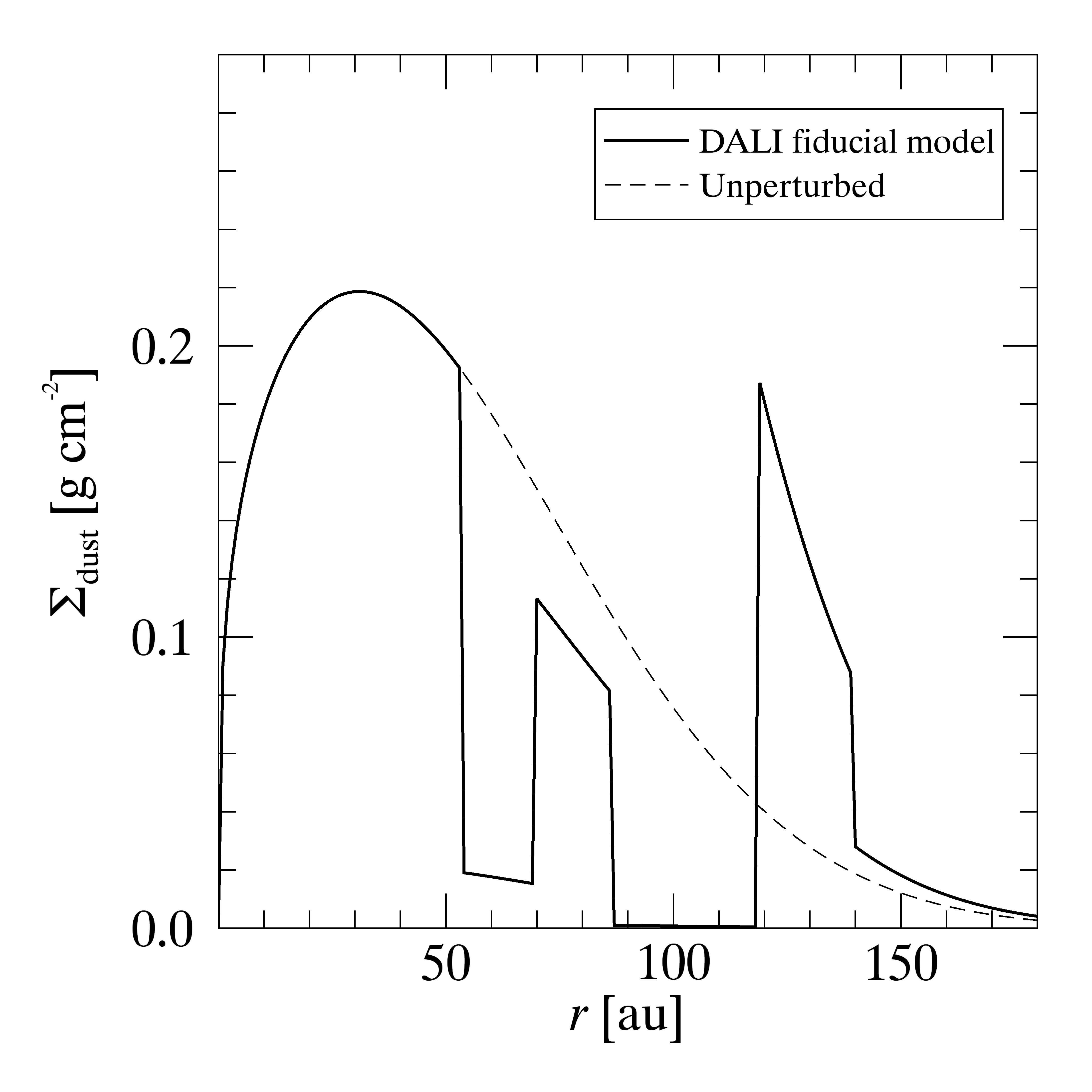}
\caption{
Dust surface density of the \textsc{dali} fiducial model. The dashed line corresponds to the initial unperturbed profile (power-law with exponential tail).}\label{fig:dali_sd}
\end{figure}

\noindent
We also fit simultaneously for the disk inclination $i$, the East-of-North position angle $PA$, and phase center offset ($\Delta \alpha$, $\Delta \delta$). 
For each set of values of the free parameters we compute a synthetic image of the model assuming axisimmetry and we use the \textsc{galario} library 
\citep{Tazzari17b} to compute the synthetic visibilities (by sampling the Fourier transform of the model image in all the observed $(u,v)$-points) 
and the resulting $\chi^2$ as 

\begin{equation}
\chi^{2} = \sum_{j=0}^{N}|V_{\mathrm{obs}}(u_{j}, v_{j})-V_{\mathrm{mod}}(u_{j}, v_{j})|^{2} w_{j}\,,
\end{equation}\label{eq:chi}

\noindent
where $w_{j}$ is the weight of the observed $(u_{j},v_{j})$ visibility point. The posterior of each model is then computed as $\exp(-\chi^2/2)$ and sampled with 80 chains 
for 45000 steps (after 5000 burn-in steps). The chains, which reached a good convergence, are shown in Fig.~\ref{fig:mcmc_chains} in Appendix~\ref{ap:mcmc}, 
in the form of marginalized 1D and 2D distributions. The parameters that we infer from the fit of the visibilities are presented in Table~\ref{tab:mcmc}: for each parameter, 
we estimate its value as the median of the marginalized distribution and its uncertainty as half the interval between 16\% and 84\% percentiles.  

\smallskip
\noindent
We find that the 1.3~mm continuum brightness distribution of AS~209 can be explained very well 
by a profile with two deep gaps ($\delta_{G1}\sim0.03$, $\delta_{G2}\sim 0.025$) at 62 and 103\,au, respectively, 
and an excess ring ($\delta_{R2} \sim 4$) at $\sim$ 130\,au. 
The excellent agreement of this profile 
with the observations is apparent in Fig.~\ref{fig:as209_mcmc}, where the deprojected 
synthetic visibilities match the observed ones up to 1500k$\lambda$.

\begin{table*}[!t]
\centering
\caption{\textsc{dali} fiducial disk model}
\begin{tabular}{llllll}
\hline
\hline
\multicolumn{6}{l}{Fixed}\\
Parameter & \multicolumn{4}{c}{Value} & Description \\
\hline
$M_{\star}$ [$M_{\odot}$]  & \multicolumn{4}{c}{0.9\tablefootmark{$\dagger$}} & stellar mass \\
T$_{\rm eff}$ [K]          & \multicolumn{4}{c}{4250\tablefootmark{$\dagger$}} & stellar temperature \\
$L_{\star}$ [$L_{\odot}$]  & \multicolumn{4}{c}{1.5\tablefootmark{$\dagger$}} & stellar luminosity \\
$d$ [pc]                           & \multicolumn{4}{c}{126} & stellar distance \\
$R_{\rm in}$ [au]          & \multicolumn{4}{c}{0.1} & disk inner radius \\
$R_{\rm c}$ [au]           & \multicolumn{4}{c}{80}   & disk critical radius \\
$R_{\rm G1}$ [au]         & \multicolumn{4}{c}{62}   & Gap 1 center \\
$hw_{\rm G1}$ [au]        & \multicolumn{4}{c}{8}    & Gap 1 half width \\
$R_{\rm G2}$ [au]         & \multicolumn{4}{c}{103}   & Gap 2 center  \\
$hw_{\rm G}$ [au]         & \multicolumn{4}{c}{16}   & Gap 2 half width \\
$R_{\rm R2, out}$ [au]     & \multicolumn{4}{c}{140}  & Ring 2 outer radius \\
$R_{\rm out}$ [au]        & \multicolumn{4}{c}{200} & disk outer radius \\
$i$ [$^{\circ}$]              & \multicolumn{4}{c}{35} & disk inclination  \\
$PA$ [$^{\circ}$]          & \multicolumn{4}{c}{86} & disk position angle \\ 
$\chi$, f$_{\rm large}$    & \multicolumn{4}{c}{0.2, 0.85} & settling parameters\\
$\psi$                   & \multicolumn{4}{c}{0.1\tablefootmark{$\dagger$}} &  flaring  exponent\\
$h_{\rm c}$            &  \multicolumn{4}{c}{0.133\tablefootmark{$\dagger$}} &  scale height at $R_{\rm c}$ \\
\hline
\multicolumn{6}{l}{Variable}\\
Parameter & Min & Max &  Step  & Fiducial & \\
\hline
$M_{\rm dust}$ [$M_{\odot}$]   & $1 \cdot 10^{-4}$ & $5 \cdot 10^{-4}$ & $0.5 \cdot 10^{-4}$ & $3.5 \cdot 10^{-4}$  & Disk dust mass\\
$\gamma_1$    & -1.0 & 1.0 & 0.1 &  0.3 & $\Sigma (r)$  power-law exponent \\
$\gamma_2$    & 1.0 & 3.0 & 0.1 &  2.0 & $\Sigma (r)$  exponential-tail exponent \\
$\tilde{\delta}_{\rm G1}$    & \multicolumn{3}{c}{(0, 0.001, 0.01, 0.05, 0.10, 0.15, 0.20)} & 0.1 & $\Sigma_{\rm dust, large}$ scale factor  in gap 1 \\
$\tilde{\delta}_{\rm R1}$    & 0.5 & 1.0 & 0.05  & 0.75  & $\Sigma_{\rm dust, large}$ scale factor in ring 1 \\
$\tilde{\delta}_{\rm G2}$    & \multicolumn{3}{c}{(0, 0.001, 0.01, 0.02, 0.03, 0.05, 0.1)} & 0.01 & $\Sigma_{\rm dust, large}$ scale factor in gap 2 \\
$\tilde{\delta}_{\rm R2}$    & 2 & 5 & 0.5 & 4.5  & $\Sigma_{\rm dust, large}$ scale factor in ring 2 \\
$\tilde{\delta}_{\rm out}$   & 1  & 3 & 0.5 & 1.5  & $\Sigma_{\rm dust, large}$ scale factor in outer disk \\
\hline\hline
\end{tabular}\label{tab:dali}
\tablefoot{
For each variable parameter we explored a range of values between a minimum (Min) and maximum (Max) in regular steps (step). In the case of $\tilde{\delta}_{\rm G1}$ and $\tilde{\delta}_{\rm G2}$ we explored non-linearly spaced values, so we list them in parenthesis. 
References: \tablefoottext{$\dagger$}{\citet{Andrews09}}.
}
\end{table*}

\section{Analysis with a physical disk model}\label{sec:analysis}
In this Section we aim to characterize the structure of AS 209 in physical terms, starting from the observed 1.3mm continuum observation. 
This step is important to estimate the drop of the dust surface density inside the two gaps. 
For this purpose we use the dust radiative transfer code implemented
into the thermo-chemical disk model \textsc{dali} (Dust and Lines, \citealt{Bruderer12, Bruderer13}). Starting from an input radiation field and from a disk density structure, \textsc{dali} solves the two 
dimensional dust continuum radiative transfer and  determines the dust temperature and radiation field strength at each disk position. 

\subsection{Model description}
We adopt the characterization of the surface brightness presented in the previous Section as a first guess for the functional form and the location of the 
gaps to be used for the surface density of the physical disk model that we use in this Section. 
The dust surface density is:

\begin{equation}\label{eq:dali_sd}
\Sigma_{\rm dust}(R) = \tilde{\delta} (R) \ \Sigma_{\rm c}  \ \Bigg(\frac{R}{R_{\rm c}}\Bigg)^{\gamma_1} \ \exp\Bigg[ - \Bigg( \frac{R}{R_{\rm c}} \Bigg)^{\gamma_2} \Bigg]
\end{equation}

\noindent
where the surface density scaling factor ($\tilde{\delta}$) is parametrized as in eq.~\ref{eq:delta}. 

\smallskip
\noindent
In the vertical direction, the density follows a Gaussian distribution with scale height $h$ ($=H/R$)

\begin{equation}
h = h_{\rm c} \Bigg( \frac{R}{R_{\rm c}}\Bigg)^{\psi} 
\end{equation}

\begin{figure*}[!t]
\centering
\includegraphics[width=18cm]{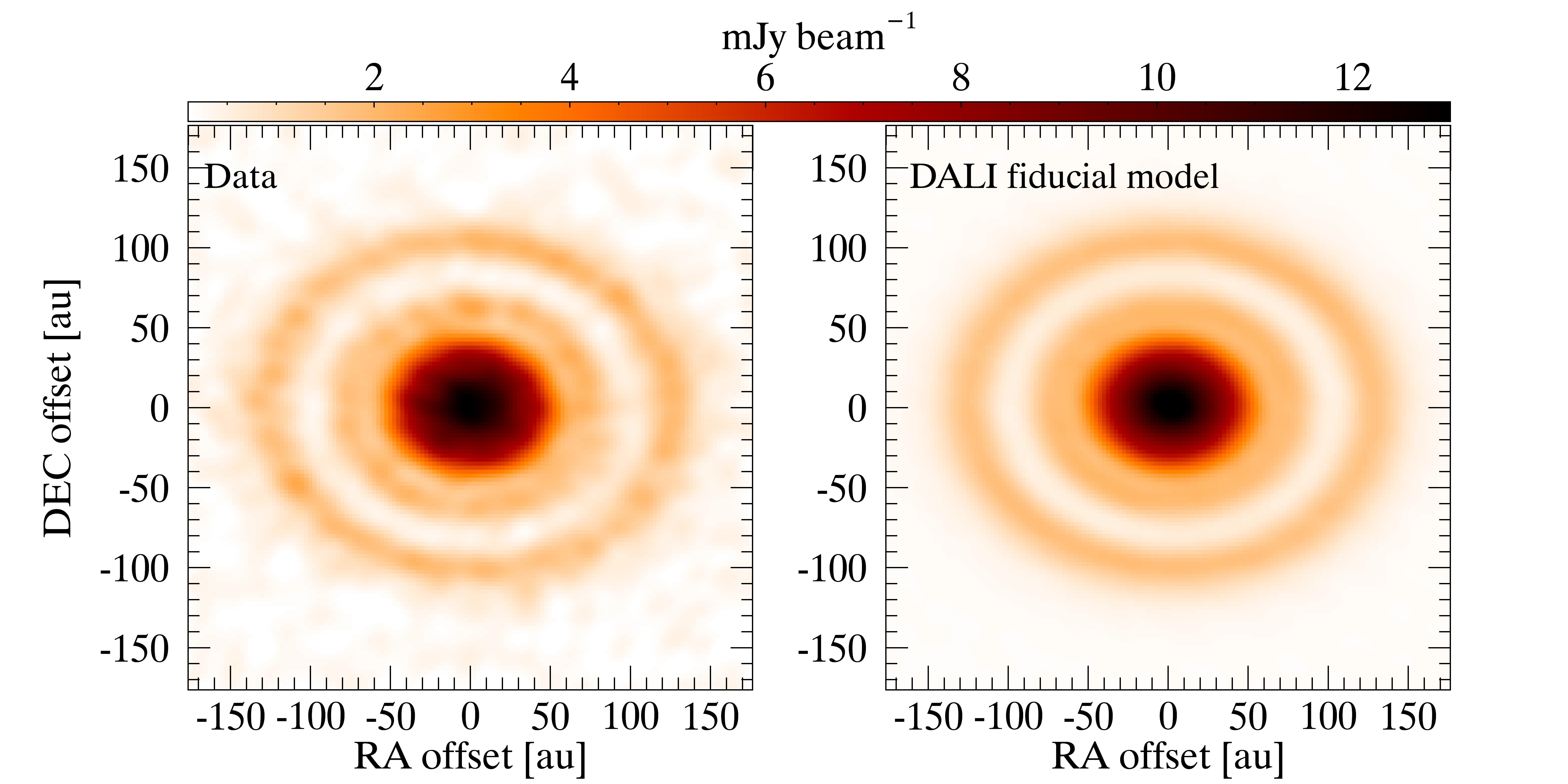}
\caption{
\textsc{dali} fiducial model, comparison of 1.3\,mm continuum image. The model image is produced with \textsc{casa.clean} starting from the synthetic 
visibilities and adopting the same \textsc{clean} parameters as the observations.  }\label{fig:dali}
\end{figure*}

\noindent
with $h_{\rm c}$ the critical scale height and $\psi$ the flaring exponent. The critical scale height ($h_{\rm c}=0.13$) and disk flaring ($\psi=0.1$) 
are taken from \citet{Andrews11}. In the adopted version of \textsc{dali}, dust settling is included following \citet{dAlessio06}, i.e., adopting two 
power-law grain size populations with different scale heights: the small grains have a scale height equal to $h$ (similar to the gas) while the 
scale height of the large grains is $\chi h$ (with $\chi < 1$) to account for the settling of the large grains. Finally, the total dust mass is distributed 
between the two populations and it is regulated by the parameter $f_{\rm large}$ (large-to-small mass ratio): thus, the dust surface density is 
$\Sigma_{\rm dust}  \cdot (1 - f_{\rm large})$ and $\Sigma_{\rm dust} \cdot f_{\rm large}$ for the small and large grains, respectively. The flaring 
parameters are fixed : $\chi = 0.2$, $f_{\rm large} = 0.85$. 

\smallskip
\noindent
We fixed the grain size populations with a small population of sizes between 0.005 and 1\,\micron ~and a large one of sizes 
between 0.005 and $s_{\rm max}\,$\micron ~with both populations sharing the same power-law exponent ($p=3.5$). The dust mass absorption 
coefficients are taken from \citet{Andrews11}. 

\smallskip
\noindent
The maximum grain size of the large dust population affects the dust opacity at millimeter wavelengths with the opacity decreasing by almost an 
order of magnitude going from $s_{\rm max}$ = 0.8\,mm to 1.0\,cm \citep[e.g.,][]{Tazzari16}. This in turn has an impact the dust temperature 
and the total dust mass. In this paper we fix $s_{\rm max} = 2000\,$\micron ~in agreement with the multi-frequency continuum analysis of 
AS 209 by \citet{Tazzari16}.

\subsection{Fiducial model}
We built a grid of \textsc{dali} disk models varying the following parameters: 

\begin{itemize}
\item[$\bullet$] total dust mass
\item[$\bullet$] surface density power-law exponent ($\gamma_1$) 
\item[$\bullet$] exponential tail exponent ($\gamma_2$) 
\item[$\bullet$] dust density scaling factors 
\end{itemize}

\noindent 
The explored range of each variable parameter is listed in Table~\ref{tab:dali}.
The values of $R_{\rm c}$, gaps centers ($R_{\rm G1}, R_{\rm G2}$) and gaps sizes (half width $hw_{\rm G1}, hw_{\rm G2}$) are taken from the best-fist results of the 
multi-components analysis.

\smallskip
\noindent
Each model is compared with the ALMA continuum observation with the aim of defining a fiducial model for the dust surface density. The comparison is 
performed in the $uv$-plane: first we compute the synthetic 1.3\,mm continuum image with \textsc{dali}, then the tool \textsc{casa.simobserve} is 
used to convert the image into synthetic visibilities at the same $uv$-positions as the observations. Finally we measure the $\chi^2$ between observation 
and model after deprojecting and binning the visibilities (bin size of 30\,k$\lambda$).

\smallskip
\noindent
The fitting procedure is performed in multiple steps: first we vary the global disk properties, i.e. $M_{\rm dust}, \gamma_1, \gamma_2$ until we find 
a good agreement with the visibilities at the shortest baselines which provides a constraint to the large scale structure. During this step the dust scaling 
factors are kept fixed: $\tilde{\delta}_{\rm G1}=0, \tilde{\delta}_{\rm G2}=0, \tilde{\delta}_{\rm R1}=1, \tilde{\delta}_{\rm R2}=1, \tilde{\delta}_{\rm out}=1$. 
In a second step, we constrain the values of the scaling factors while keeping fixed $M_{\rm dust}, \gamma_1, \gamma_2$. The process is repeated
until convergence. This allows us to refine the grid resolution iteratively.   

\noindent
The fiducial model is defined by the set of parameters that minimize the $\chi^2$ between the observed and synthetic visibilities (eq.~\ref{eq:chi}) 
within the explored parameter space.

\smallskip
\noindent
The parameters of the \textsc{dali} fiducial model are listed in Table~\ref{tab:dali}. Figure~\ref{fig:dali_sd} shows the dust surface density of 
the fiducial model and the model image is shown in Figure~\ref{fig:dali}. We note that, in order to quantitatively reproduce the disk structure 
(gaps and rings) with DALI we need to set different dust scaling factors for the two gaps: $\tilde{\delta}_{\rm G1} \sim 0.1$ and $\tilde{\delta}_{\rm G2} \sim 0.01$. 

\begin{figure*}[!t]
\centering
\includegraphics[width=18cm]{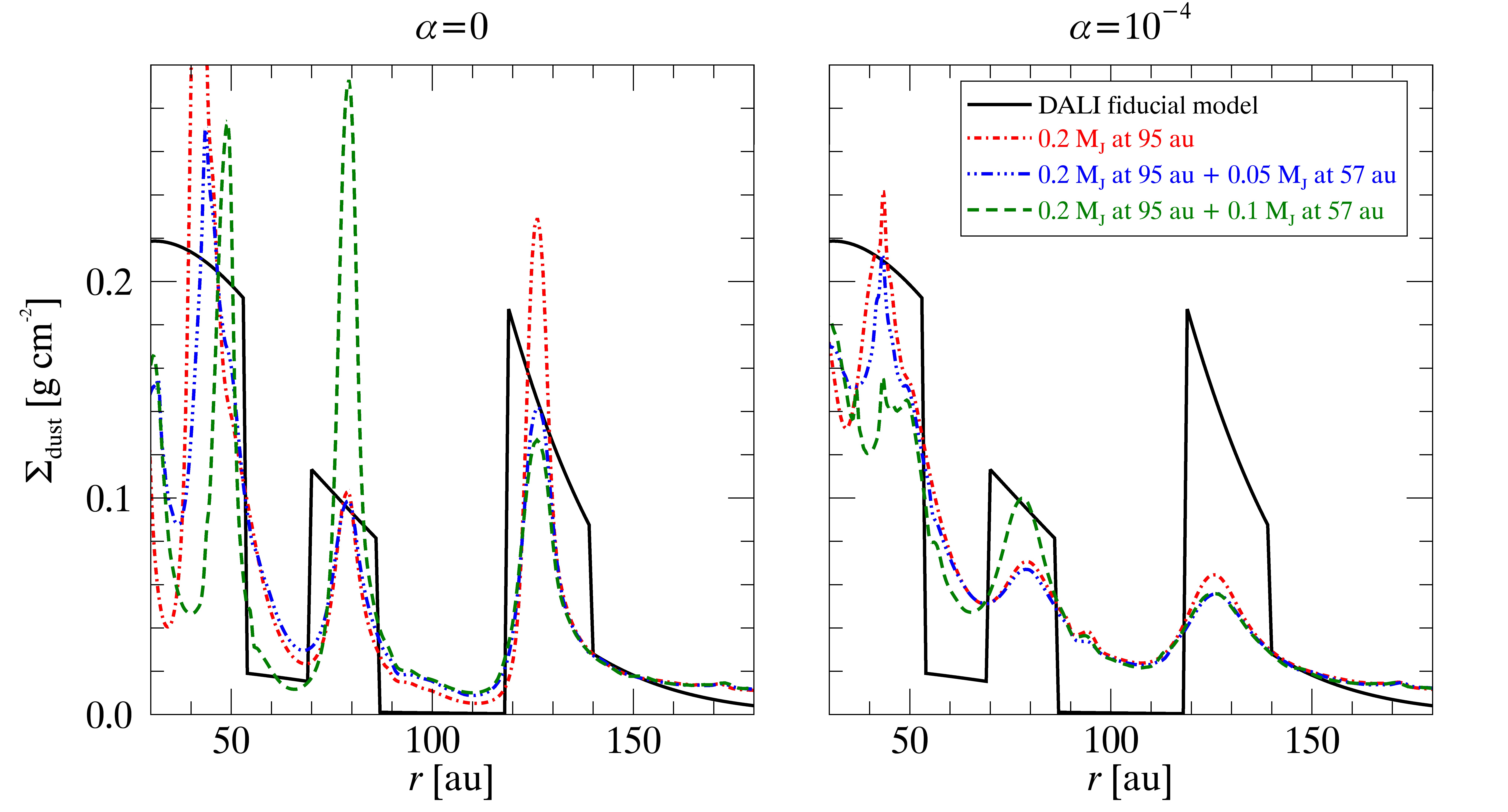}
\caption{Results of hydrodynamical simulations for the inviscid (left) and $\alpha=10^{-4}$ (right) case. The black line corresponds to the dust surface
density determined with \textsc{dali} (Sect.~\ref{sec:analysis}). The (red) dot-dashed line is the 1\,mm grains surface density based on the single planet
scenario with 0.2\,M$_{\rm J}$ planet at 95\,au injected after 0.65 Myr (Sect.\ref{sec:hydro}). The (blue) dot-dot-dashed and the (green) dashed curves
corresponds to the two planets scenario with an inner planet of mass 0.05\,M$_{\rm J}$ and 0.1\,M$_{\rm J}$, respectively.}
\label{fig:hydro}
\end{figure*}

\section{Comparison to hydrodynamical simulations}\label{sec:hydro} 
In this section we investigate the possibility that the gaps were produced by planets, as has been suggested for other disks, e.g. HL~Tau, TW Hydra, 
HD 163296, and HD 169142 \citep{alma15, Andrews16, Isella16, Fedele17}.   To do this, we have run 2D simulations of planet-disk interaction using a 
version of FARGO-3D \citep{Benitez16} modified to include dust dynamics \citep{Rosotti16}. The disk parameters were chosen to match the best fit model 
(see Table \ref{Tab:SimParams}). The simulations were run using a logarithmic radial grid, extending from 10~au to 300~au at a resolution of 
$Nr \times N\phi = 550 \times 1024$. Given the relative expense of the hydrodynamic simulations, we have not conducted an exhaustive fit to the data, 
instead we investigate the typical planetary properties and disk conditions that produce reasonable gap structures. The key information available for 
constraining the planets properties are the gap width, depth and location. The position of the two gap centers (62 and 103\,au) are close enough to 
the ratio of radii expected for two planets migrating together in 2:1 resonance (with semi-major axis ratio 0.63), which forms the starting point 
for our investigation. 

\smallskip
\noindent
Starting from the relationship between the width of a gap opened by a planet and its mass derived by \citet{Rosotti16}, we already see that the inner 
planet must be low mass due because the inner gap is narrow (just 16\,au or approximately 2 -- 3 pressure scale-heights). This suggests that the mass 
of the inner planet is in the Neptune mass regime ($\sim 15\,M_\oplus$, i.e. $0.05\,M_\mathrm{J}$). 
However, while \citet{Rosotti16} showed that these planets can produce observable features, they found gap 
depths much smaller than the inner gap in AS~209. Nonetheless, the depth of the gap is senstive to disk 
viscosity, with planets opening deeper gaps in low viscosity disks \citep{Crida06,Zhu13}. Thus together 
with the gap width, the gap depth places constraints on both the planet mass and disk viscosity. 
Furthermore, at extremely low viscosity, \citet{Dong17} and \citet{Bae17} showed that planets may open multiple dust gaps 
in inviscid disks, which raises the interesting possibility that the gaps in AS~209 may be opened by a 
single planet.

\smallskip
\noindent
From the right panel of Fig.~\ref{fig:hydro}, we see that even a modest viscosity $\alpha = 10^{-4}$ is too high to produce 
deep enough gaps to explain the structures in AS~209.
While a single $0.2\,M_\mathrm{J}$ planet at 95\,au matches the  width of the outer gap, the drop in dust 
surface density it is far too shallow compared to what is inferred. Although more massive planets can open deep enough gaps 
(for $\alpha = 10^{-4}$), they produce structures that are too wide and begin to prevent the inflow of 
dust entirely. This suggests that the outer gap is consistent with a $0.2\,M_\mathrm{J}$ planet, but 
requires even lower viscosity.

\smallskip
\noindent
The inviscid ($\alpha = 0$) simulations produce a much better match to the inferred gap structure 
(Fig.~\ref{fig:hydro}, left panel). Already a single $0.2\,M_\mathrm{J}$ planet at 95~au is in 
remarkable agreement with the gap structure, producing a deep outer gap along with a second inner 
gap at approximately the the right location and with a reasonable width and depth. 
The depth of the gaps is not only sensitive to viscosity, but in the inviscid case it also increases 
with time, thus being a much weaker tracer of the planet mass than the position of the peaks (as noted by 
\citealt{Rosotti16}). However, the width of the gaps are similar in both the viscous and inviscid 
cases. Interestingly, the simulations produce an additional gap at around 35\,au, which coincides 
with similar, but much smaller amplitude, feature in the radial intensity profile 
(Fig.~\ref{fig:radial}). Given that the depth of this feature is dependent on parameters such as 
viscosity, and optical depth effects may further reduce the amplitude of variation in the observed 
intensity profile, it is possible that this inner structure is related to the presence of a planet 
near 100\,au. 

\smallskip
\noindent
Since low viscosity is required to reproduce the gap structure in AS~209, and in this case a single 
planet may produce both gaps, it is interesting to consider whether there still could be a second 
planet present. Thus we have re-run both the viscous and inviscid simulations with an inner planets 
at 57~au, close to the 2:1 resonance with the outer planet. In both cases the planet mass is consistent 
with the estimate from the relationship between gap width and planet mass: a $0.05\,M_\mathrm{J}$ planet 
produces a negligible difference to the structure of the gaps and could thus be easily hidden in the gap. 
The largest planet mass compatible with the gap widths is roughly $0.1\,M_\mathrm{J}$, with larger planet 
masses producing a gap that is too wide in the inviscid case. While a slightly more massive planet may 
be compatible with the inner gap when $\alpha = 10^{-4}$, this is hard to reconcile with the need for a 
lower $\alpha$ in the outer gap.

\begin{table}
\centering
\caption{Disk model used in the planet disk-interaction simulations}
\label{Tab:SimParams}
\begin{tabular}{ll}
\hline\hline
Parameter & Value \\
\hline
Temperature & $190 (R/100\,\mathrm{au})^{-0.55}$ \\
Gas Surface Density &$ 7.5 (R/100\,\mathrm{au})^{-1}$ \\
Grain size & 1~mm \\
Viscous $\alpha$ & 0, $10^{-4}$ \\
\hline\hline
\end{tabular}
\end{table}

\section{Conclusions}\label{sec:conclusion} 
The ringed dust structure of AS 209 revealed by ALMA is consistent with the presence of a Saturn-like 
($M_{\rm planet} = 0.2 M_{\rm J} = 0.67\,M_{\rm Saturn}$) planet at $r = 95\,$au. The planetary mass is 
constrained by the width and depth of the dust gap: for the chosen gas properties, our hydrodynamical 
simulations indicate that less massive planets ($\lesssim 0.5 \,M_\mathrm{Saturn}$) do not produce a gap 
that is sufficiently wide, while more massive planets ($\gtrsim 1.0 \,M_\mathrm{Saturn}$) prevent the 
transport of dust inwards entirely, forming a transition disk (inner hole in large grains). 
Our radiative transfer calculations (Sect~\ref{sec:analysis}) show that surface density of the mm grains 
beyond the planet orbit is enhanced and largely confined in a narrow ring. This may be the signature of 
dust pile-up due to the planet-induced gas pressure maximum beyond the orbit of the planet. Interestingly, 
the ALMA C$^{18}$O image presented in \citet{Huang16} shows an extended emission peaking at nearly 130\,au. 
This extended emission is co-spatial to the outer dust ring. This is a strong indication that the large 
scale C$^{18}$O emission follows the actual disk surface density in the outer disk and it points to a gas 
density drop by a factor of a few inside the outer gap (`$G2$'). 

\smallskip
\noindent
We have also investigated the existence of a second planet in correspondence of the inner dust gap. 
We conclude that there could be a second planet present in the inner gap if it is less than about 
$0.1\,M_\mathrm{J}$. Since the two gaps are close to the 2:1 resonance this raises the possibility 
that the structure could be caused by a pair of planets migrating in resonance. The 1\,mm surface
density resulting from a pair of 0.05\,$M_{\rm J}$ -  0.2\,$M_{\rm J}$ is in remarkable agreement
with the observations. Nonetheless, while both scenarios require low disk turbulence, the presence of 
the inner planet is not needed to explain the observed structures.

\smallskip
\noindent
The inferred presence of the Saturn-like planet at $r \sim 95\,$au raises 
new questions about planet formation at such large distance from the star.
Gravitational instability can occur on short timescales and is a 
viable process for the formation of giant planets on wide orbits \citep[e.g,][]{Kratter16}.
An alternative scenario is pebble accretion, in which planets grow through the accretion of cm-sized (or larger) 
grains that are weakly coupled to the gas \citep{Lambrechts12}. In this case however planet formation is challenged 
by dust migration: according to dust evolution models such large dust grains are expected to drift towards the star 
in much less than a Myr \citep[e.g.,][]{Takeuchi05, Brauer08}, which is difficult to reconcile with disk lifetimes 
(a few Myr, e.g., \citealt{Fedele10}). 

\smallskip
\noindent
In the case of AS~209, the dust inward migration is likely to be slowed down by the presence of radial 
dust traps induced by the presence of a Saturn-like planet. While this can help to reconcile the presence 
of large dust with the disk's lifetime, it must hinder pebble accretion in the inner disk, restricting 
this mode of planet formation to the earliest phases of disk evolution, on timescales $\lesssim 1\,$Myr. 

\begin{acknowledgements}
This paper makes use of the following ALMA data: ADS$/$JAO.ALMA$\#$2015.1.00486.S. 
ALMA is a partnership of ESO (representing its member states), NSF (USA) and NINS
(Japan), together with NRC (Canada), NSC and ASIAA (Taiwan), and KASI (Republic 
of Korea), in cooperation with the Republic of Chile. The Joint ALMA Observatory 
is operated by ESO, AUI$/$NRAO and NAOJ.
DF acknowledges support from the Italian Ministry of Education, Universities and 
Research, project SIR (RBSI14ZRHR). 
MT has been supported by the DISCSIM project, grant agreement 341137 funded by the European Research Council under ERC-2013-ADG.
\end{acknowledgements}

\begin{appendix}

\section{MCMC}\label{ap:mcmc}

\begin{figure}[!t]
\begin{center}
\includegraphics[width=9cm]{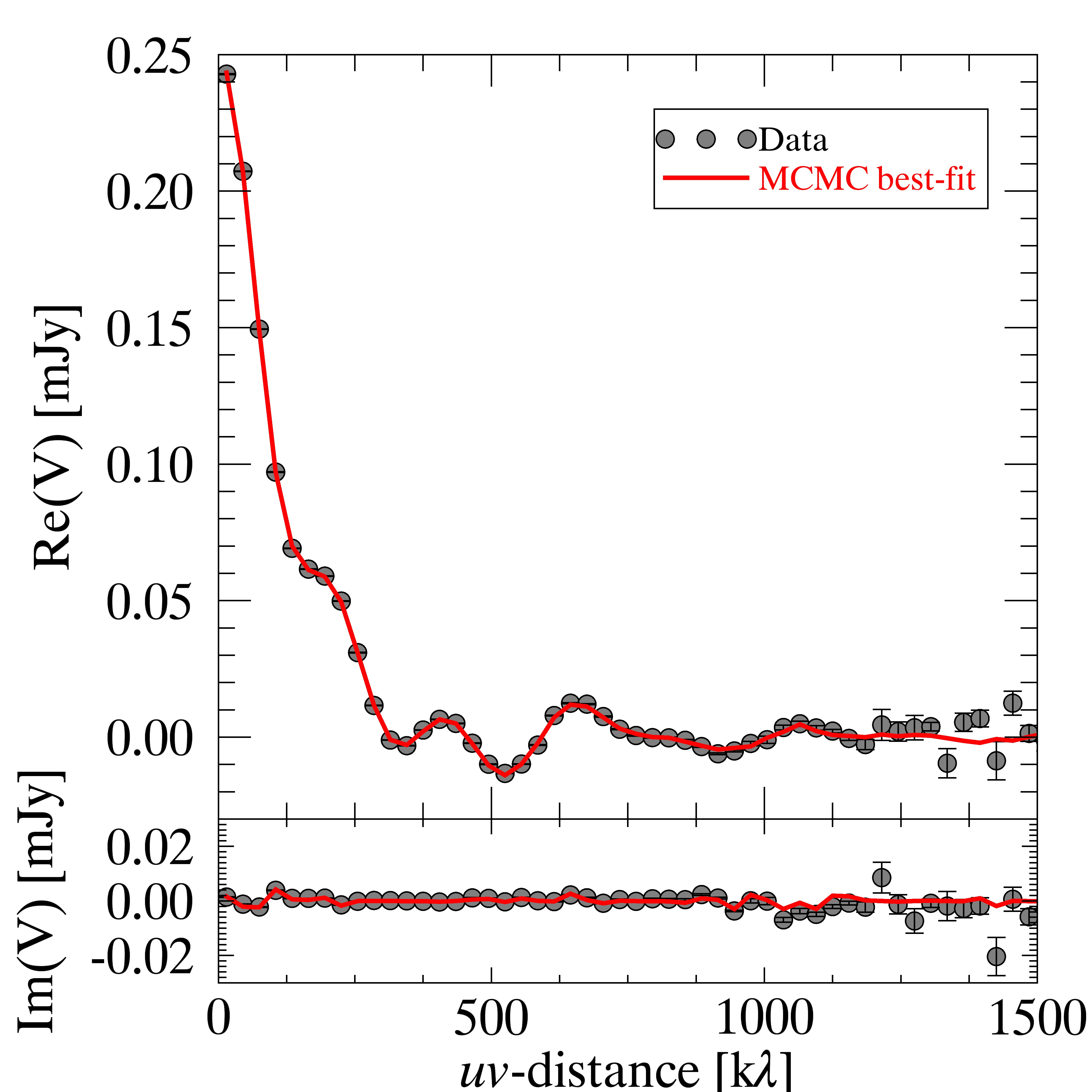}
\caption{ Results of multi-components geometrical fit with MCMC: comparison of the observed (gray dots) and bestfit model (red line) deprojected visibilities.}
\label{fig:as209_mcmc}
\end{center}
\end{figure}

\begin{figure*}[htbp]
\begin{center}
\includegraphics[width=18cm]{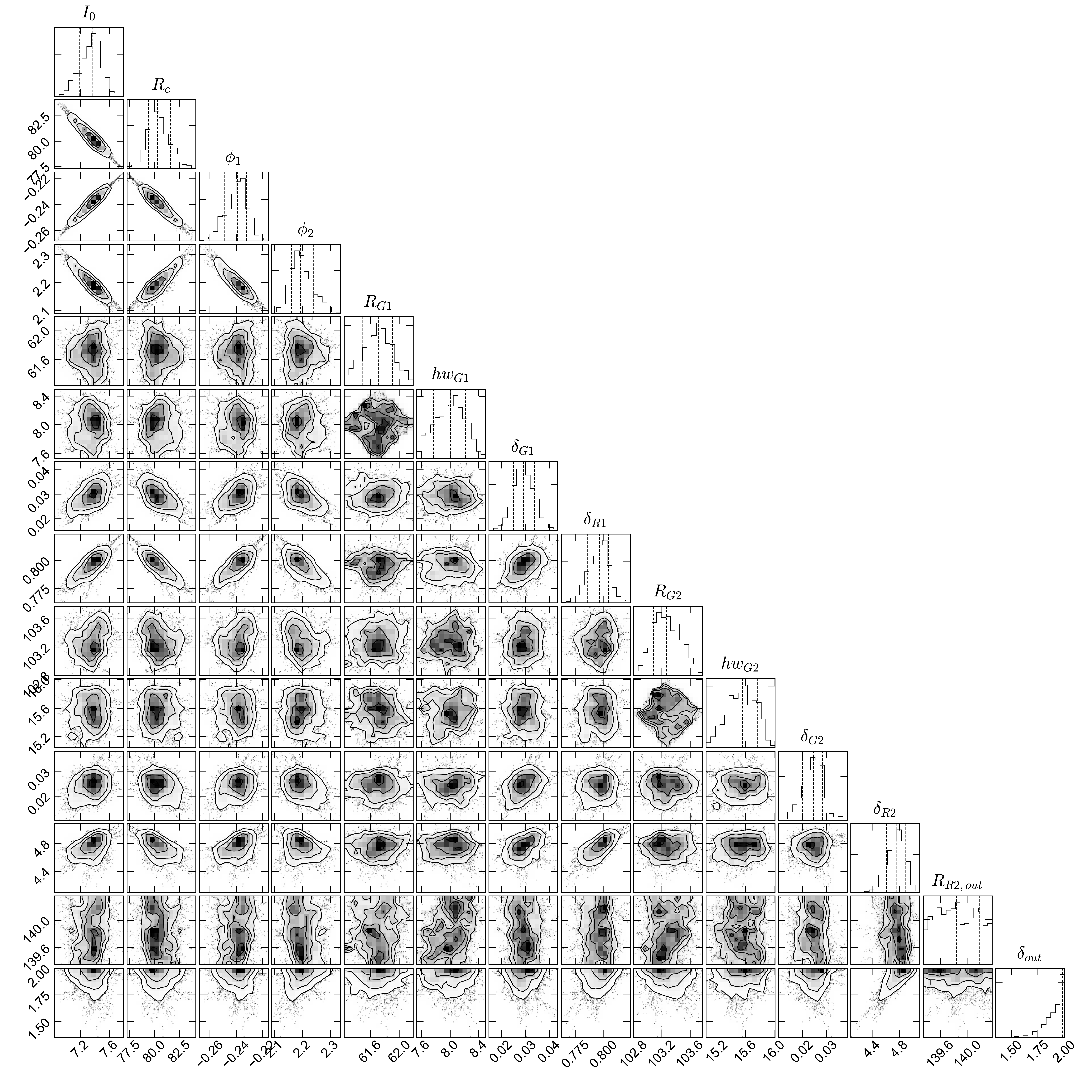}
\caption{1D and 2D marginalized distributions of the posterior sampling obtained from MCMC. In the 1D marginalized distributions of each parameter the vertical dashed lines represent the 16th, 50th and 84th percentiles: the median is taken as the bestfit value, and half the interval between 16th and 84th percentiles as the uncertainty.} 
\label{fig:mcmc_chains}
\end{center}
\end{figure*}

\end{appendix}

\end{document}